\documentclass[a4paper,11pt]{article}
\pdfoutput=1
\usepackage{jcappub}
\usepackage{amsthm,graphicx,multirow}
\usepackage{epsfig}
\usepackage{latexsym, amssymb} 
\usepackage{amsmath}
\usepackage{epstopdf}
\usepackage{xcolor}
\def\beq{\begin{equation}}
\def\eeq{\end{equation}}
\def\br{\begin{eqnarray}}
\def\er{\end{eqnarray}}
\def\benu{\begin{enumerate}}
\def\efnu{\end{enumerate}}

\begin{document}
\title{Reionization in the dark and the light from Cosmic Microwave Background}
\author[1,2]{Dhiraj Kumar Hazra,}
\author[1,2]{Daniela Paoletti,}
\author[1,2]{Fabio Finelli,}
\author[3,4,5,6]{George F. Smoot}
\affiliation[1]{Istituto Nazionale Di Fisica Nucleare, Sezione di Bologna,Viale Berti Pichat, 6/2, I-40127 Bologna, Italy}
\affiliation[2]{Osservatorio di Astrofisica e Scienza dello Spazio di Bologna/Istituto Nazionale di Astrofisica, via Gobetti 101, I-40129 Bologna, Italy}
\affiliation[3]{AstroParticule et Cosmologie (APC)/Paris Centre for Cosmological Physics, Universit\'e
Paris Diderot, CNRS, CEA, Observatoire de Paris, Sorbonne Paris Cit\'e University, 10, rue Alice Domon et Leonie Duquet, 75205 Paris Cedex 13, France}
\affiliation[4]{Institute for Advanced Study \& Physics Department, Hong Kong University of Science and Technology, Clear Water Bay, Kowloon, Hong Kong}
\affiliation[5]{Physics Department and Lawrence Berkeley National Laboratory, University of California, Berkeley, CA 94720, USA}
\affiliation[6]{Energetic Cosmos Laboratory, Nazarbayev University, Astana, Kazakhstan}
\emailAdd{hazra@bo.infn.it, paoletti@iasfbo.inaf.it, finelli@iasfbo.inaf.it, gfsmoot@lbl.gov} 

\abstract 
{We explore the constraints on the history of reionization from Planck 2015 Cosmic Microwave Background (CMB) data and we derive the 
forecasts for future CMB observations. We consider a class of monotonic histories of reionization as 
parametrized by two additional extra parameters with respect to the average optical depth used in the 
instantaneous reionization modeling. We investigate the degeneracies between the history of reionization and selected extensions of
the standard cosmological model. In particular, we consider the degeneracies with the total mass of the neutrino sector and we 
discuss the possible correlation between the dark matter annihilation and the duration of reionization in the CMB. We use an extension 
to poly-reion model that was proposed in~\cite{Hazra:2017gtx}. We compare the constraints from Planck 2015 data with the predicted 
constraints from possible future CMB mission as LiteBIRD, and we also use the proposed CORE-like specifications as an example of what 
higher resolution can bring in addition. We find that the degeneracy between the averaged optical depth and the duration of reionization
will be substantially removed by both concepts. Degeneracies between the reionization history and either the total neutrino mass and 
properties of dark matter annihilation will also be improved by future surveys. We find only marginal improvement in the constraints 
on reionization history for the higher resolution in the case of long duration of reionization.}

  \maketitle

\section{Introduction}
CMB anisotropies are one of the fundamental cosmological observables which may shed light on the early stages and following evolution of the Universe. 
The ever increasing accuracy of observational instruments has already allowed cosmic variance and foreground limited observations of temperature 
anisotropies up to high multipoles, and the future experiments try to achieve the same with polarization anisotropies. CMB anisotropies in polarization 
are crucial to investigate the epoch of reionization, current measurements on large angular scales are still affected by systematics but future
experiments promise to reach the cosmic variance limit in the determination of the optical depth. 
This would allow constraining reionization models and to break their degeneracies with extensions of the standard cosmological model.
In fact, apart from the primordial physics, that we have considered recently in~\cite{Hazra:2017joc}, other physical processes can also be degenerate with
the process of reionization and therefore can be of hindrance in understanding the reionization history with CMB. 

For example if we allow the neutrino mass to vary, Planck 2015 data (temperature, polarization and lensing) prefer a higher value of optical depth 
($\tau_{\rm mean}=0.074$) compared to the baseline case where the total neutrino mass is fixed to be $0.06~{\rm eV}$ ($\tau_{\rm mean}=0.063$). 
This comparison is provided when we assume ${\rm Tanh}$ model of reionization. We explore this degeneracy using extended reionization histories,
in particular, using a modified form of poly-reion model. 
In this construct we also address the degeneracy between the duration of reionization and the average optical depth. 

In a cosmological scenario where dark matter annihilates, this process increases the free electron fraction by heating up the inter-galactic medium (IGM), 
and can be marginally degenerate with the reionization process. We present the constraints on dark matter annihilation and investigate its degeneracy with 
the duration of reionization. We begin by using the Planck 2015 constraints to explore the current degeneracies, then we use the  allowed region of parameter
space to derive fiducial angular power spectra and compare them in the light of future space based CMB mission proposal LiteBIRD. We also provide a comparison 
with projected constraints from a concept with higher angular resolution, taking as an example, the specifications of CORE~\cite{core:inst,Delabrouille:2017rct}. 

This paper is organized as follows: in the next section, we discuss the model of reionization that constructs the reionization history. 
In sec.~\ref{sec:data} we discuss the datasets that have been used in the analysis and survey configurations for the future missions. 
In sec.~\ref{sec:degen} we present the possible degeneracies between reionization and other physical processes that we explore. Following that, 
in sec.~\ref{sec:results} we provide present constraints and futuristic forecasts. Finally we conclude in sec.~\ref{sec:conclusion}.

\section{The reionization history}
We use the poly-reion model of reionization with a simple extension. In the original model~\cite{Hazra:2017gtx} two of us have introduced a smooth history of reionization using nodes fixed at different redshifts and connecting them with Piecewise Cubic Hermite Interpolating Polynomial (PCHIP). Since the present available data does not provide tight constraints on the large scale polarization spectrum, their ability to constrain the detailed history of reionization is rather limited. In this paper, since we consider future CMB observations, we allow a node to vary in the poly-reion formalism. The free electron fraction ($x_e$) at a redshift $z$ is given by:
\begin{equation}
x_e(z)=(1+F_{\rm He})f(z)~\label{eq:polyreion},
\end{equation}
where $f(z)$ is the polynomial and the contribution to the $x_e$ from the first ionization of Helium is given by $F_{\rm He}$. Following different observations of Lyman-$\alpha$ 
forest spectrum~\cite{Bouwens:2015vha,Fan:2005es} we impose that reionization is complete by $z=5.5$. However we do not make any assumption regarding the beginning of reionization
as in~\cite{Hazra:2017joc} and we denote it as $z_{xe=0}$. In between this two nodes, in this paper, we just consider one moving node at internal redshift $z_{\rm int}$ and we 
define the $f(z)$ at that redshift to be $x_e^{\rm H}(z_{\rm int})$. The entire history of reionization is then constructed using PCHIP through these nodes. While this variation 
of poly-reion has 3 free parameters, namely $z_{\rm int}$, $x_e^{\rm H}(z_{\rm int})$ and $z_{xe=0}$, in our analysis, instead of using $z_{xe=0}$ as free parameter, we use the 
optical depth $\tau$ as free parameter for faster convergence. Similar to~\cite{Hazra:2017joc} we solve for $z_{xe=0}$ given $\tau$ and the other 2 free parameters. We make use 
of the definition: $\tau=\int \sigma_{\rm T} n_e(z) dl$~\footnote{free electron density is $n_e(z)$ and Thomson scattering cross section is given by $\sigma_{\rm T}$}. We allow
$z_{\rm int}$ to vary between $5.5$ and $30$. We chose the redshift 30 as the upper limit as the reionization may not have started before~\cite{firstsources}. However the node at $z_{xe=0}$ allows the possibility 
of having some electron fraction even before redshift 30, apart from the residual fraction since recombination. Note that one node restricts our model to only allow for the monotonic
increase in electron fraction with time. 
A variation of poly-reion discussed in~\cite{Hazra:2017gtx} (conservative case) and also a simple modification of poly-reion that recently appeared in~\cite{Millea:2018bko} allow for
oscillations in the $x_e$, however we restrict here to the minimal monotonic case described above. 

\section{Present datasets, proposed missions and priors}~\label{sec:data}
We consider the currently available Planck 2015 data {\it i.e.} Planck high-$\ell$ Plik TT,TE,EE likelihood in combination with low-$\ell$ 
joint temperature and polarization likelihood based on the Commander component separated map in temperature and the LFI 70 GHz data in 
polarization (cleaned with the 30 GHz and 353 GHz for syncrotron and dust contamination respectively) and the Planck lensing likelihood. 
There has been a further update to the low-$\ell$ polarization likelihood  
based on HFI cross spectra~\cite{Aghanim:2016yuo,Adam:2016hgk} which results in a lower value of the optical depth with smaller errors. 
However this likelihood is not publicly available and we therefore 
consider only Planck 2015 data. 

As an example of future high sensitivity polarization experiment we have chosen the LiteBIRD~\cite{Matsumura:2013aja} satellite, a large scale polarization 
dedicated proposed mission to JAXA actually in phase-A study. 
We use the specifications provided in~\cite{Matsumura:2016sri}. We use the temperature and E-mode polarization up to $\ell=1350$. The instrumental 
specifications for the central range of frequencies are:
\begin{eqnarray}
{\mathrm{Frequency\, [GHz]}}&=& \{78.0,88.5,100,118.9,140,166,195\} \\
\mathrm{FWHM\, [Arcmin]}&=&\Big\{55,49,43,36,31,26,22 \Big\}\nonumber \\
\mathrm{\Delta T\, [\mu K\,arcmin]}&=&\Big\{10.82,8.77,11.03,8.91,5.87,6.15,4.74\Big\}\nonumber \\
\mathrm{\Delta P\, [\mu K\,arcmin]}&=&\Big\{15.3,12.4,15.6,12.6,8.3,8.7,6.7\Big\}\nonumber \\
\end{eqnarray}

As an example of higher resolution concept for a CMB space mission dedicated to CMB polarization we have considered the specifications of the CORE proposal for the M5 call to 
ESA~\cite{Delabrouille:2017rct}. For the range of frequencies 130-220 GHz the specifications are:

\begin{eqnarray}
{\mathrm{Frequency\, [GHz]}}&=& \{100,115,130,145,160,175,195,220\} \\
\mathrm{FWHM\, [Arcmin]}&=&\Big\{8.51,7.68,7.01,6.45,5.84,5.23 \Big\}\nonumber \\
\mathrm{\Delta T\, [\mu K\,arcmin]}&=&\Big\{3.9,3.6,3.7,3.6,3.5,3.8\Big\}\nonumber \\
\mathrm{\Delta P\, [\mu K\,arcmin]}&=&\Big\{5.5,5.1,5.2,5.1,4.9,5.4\Big\}\nonumber \\
\end{eqnarray}

As in the CORE science forecast  papers \cite{core:inf} \citep{core:cosmoparam} we use idealized inverse-Wishart likelihood, 
we do not take into account possible systematics and we use a mitigation of foreground residuals.
Following what is done in CORE science papers we assume that lower and higher frequency channels of LiteBIRD and CORE suffice to 
remove foreground contamination. Ref.~\cite{Remazeilles:2017szm} demonstrates that this is a reasonable approximation
for E-mode polarization.

\section{Reionization history and its degeneracy with cosmological parameters}~\label{sec:degen}
The CMB photons carry convolved signals from the primordial perturbations and their evolution till today including the footprints of the history of reionization.
Because of this interplay there is a very well known degeneracy between the optical depth and the amplitude of scalar fluctuations, an increase of the optical
depth suppresses the CMB anisotropy peaks leading to a degeneracy  which involves to a some extent the spectral index as well. It is therefore crucial to take
into account the changes induced on the E-mode polarization on large scales by a general enough reionization history.

Compared to near instantaneous $\rm Tanh$ reionization, an extended reionization history flattens the sharp dip at the large scales ($\ell \sim 15$) 
increasing the reionization bump in EE at small multipoles.
This increase is in agreement with the Planck 2015 polarization data resulting in an improvement in fit to the data with extended reionization 
models~\cite{Hazra:2017gtx,Heinrich:2018btc,Villanueva-Domingo:2017ahx,Obied:2018qdr}
compared to near-instantaneous model. Note that such type of signal in polarization is degenerate with the large scale features in the 
scalar power spectra. In this aspect, we had already considered the degeneracy with the non-standard inflationary dynamics that affects large 
scale temperature and polarization in~\cite{Hazra:2017joc}. Note that initially such degeneracy for Planck and ideal experiment was forecast
in~\cite{mortonson} and for a recent paper also see~\cite{Obied:2018qdr}. However, in~\cite{Hazra:2017joc} we found that there are no 
substantial degeneracies if we restrict ourselves to monotonic reionization histories and forecast with a fiducial primordial power spectrum with features.  

In this paper we further consider degeneracies with the history of reionization and with the dark particle sector.
First, the reionization optical depth is a line of sight integral of the free electron density. Therefore the duration of reionization is directly 
related to the optical depth. Since the temperature anisotropy spectrum has little power in constraining the optical depth by its own we expect that
there exist a substantial degeneracy if accurate large scale polarization data is not used. In this paper we define the duration of 
reionization ($\Delta_z^{\rm Reion}$) as the redshift difference between the $x_e=0.1$ and $x_e=0.99$, {\it i.e.} the difference between 10\% and 99\%
completion of reionization. Note that this definition is the same as has been used in CORE cosmological parameters analysis~\cite{core:cosmoparam}. 
The duration of reionization $\Delta_z^{\rm Reion}$ is a useful quantity as it can be constrained using CMB, even with the current data. The weak
constraint on $z_{xe=0}$ derived in~\cite{Hazra:2017joc} shows that with only CMB it is not possible to probe the beginning of reionization.

Second, we consider the degeneracy with the neutrino mass. Total mass of neutrinos affects the CMB angular power spectrum in two ways. It changes 
the large scale spectrum by contributing to the early integrated Sachs–Wolfe effect. After the photon decoupling massive neutrinos contribute to the
expansion history and thereby changes the distance measure. As we have discussed in the introduction, massive neutrinos and the optical depth of the
reionization are positively correlated. Therefore when we consider poly-reion to define the reionization history it is expected that duration of 
reionization will also be degenerate with neutrino mass. We asses the current level of degeneracy with Planck 2015 data and we forecast the capabilities
of LiteBIRD and CORE to possibly remove it. We consider only degenerate hierarchy of neutrino mass since the normal and inverted hierarchy will not 
be distinguished at more than one standard deviation ($\sigma(\Sigma m_\nu)\simeq 37-51~{\rm meV}$) even with a sensitivity like CORE. The higher 
resolution of an experiment like CORE will improve the constraints in both the reionization and neutrino mass direction, whereas coarser resolution
of LiteBIRD will allow an improvement mainly in the reionization part.

Finally we discuss the degeneracy with dark matter annihilation. The possibility of dark matter annihilation has been discussed extensively in the 
literature~\cite{Galli,Lopez-Honorez:2013lcm,Planck:2015Param,Kawasaki:2015peu,core:cosmoparam}. The annihilation of dark matter would have mainly two types of effect on the CMB angular power spectra. The reduced number of dark matter particles caused by the annihilation process will slightly impact the recombination epoch and have an effect on the small angular scales. 
The increased number of photons injected in the plasma by the annihilation induces an heating of the matter which modifies the number of free electrons. This has a strong impact on the polarization anisotropies on the intermediate scales corresponding to the decoupling time where the effect of the annihilation is stronger. This latter effect is expected to be degenerate to some extent with the reionization history that may also affect the low multipoles. CMB angular power spectrum is in particular sensitive to the parameter 
$p_{\rm ann}=\frac{f_{z=600} <\sigma v>}{m_\chi}$. Here, ${m_\chi}$ is the mass of the dark matter particles and $<\sigma v>$ represents the velocity weighted cross-section. $f_{z=600}$ is the efficiency factor taking into account the fraction of dark matter energy deposited into the plasma. In the literature this combined parameter is commonly known as $p_{\rm ann}$ with dimensions of ${\rm cm}^3/s/{\rm GeV}$. In this paper we investigate the degeneracy between the $p_{\rm ann}$ and the history of reionization.


\section{Results}~\label{sec:results}
We have used the public Einstein-Boltzmann code {\tt CAMB}~\cite{cambsite,Lewis:1999bs} to generate the predictions for the angular power spectra. 
We implemented the variation of poly-reion in {\tt CAMB}. We also use {\tt CosmoRec}~\cite{cosmorec} when we take into account the dark matter annihilation.
We have used the {\tt CosmoMC}~\cite{cosmomcsite,Lewis:2002ah} code connected to the poly-reion extension of  {\tt CAMB} we developed in order to compute 
the Bayesian probability distribution of cosmological parameters.
We vary the baryon density $\omega_{b}=\Omega_{b} h^2$, the cold dark matter density 
$\omega_{c}= \Omega_{c}h^2$ (with $h=H_0/100\, {\mathrm {km}}\,{\mathrm {s}}^{-1}{\mathrm {Mpc}}^{-1}$), the ratio of the sound horizon to the angular
diameter distance at decoupling $\theta$, $\ln (10^{10} A_S)$ (primordial spectral amplitude), $n_S$ (primordial spectral tilt), the parameters $z_{\rm int}$, $x_e^{\rm H}(z_{\rm int})$ and the optical depth 
$\tau$ for the extended reionization defined within the poly-reion model and the parameters for the extended models ($\Sigma m_\nu$ and $p_{\rm ann}$ separately). 
For the Planck 2015 data we also vary the nuisance parameters as provided in the Planck likelihood package~\cite{Planck:2015Like}.

\subsection{Present constraints}

\begin{figure*}[!htb]
\begin{center} 
\resizebox{205pt}{160pt}{\includegraphics{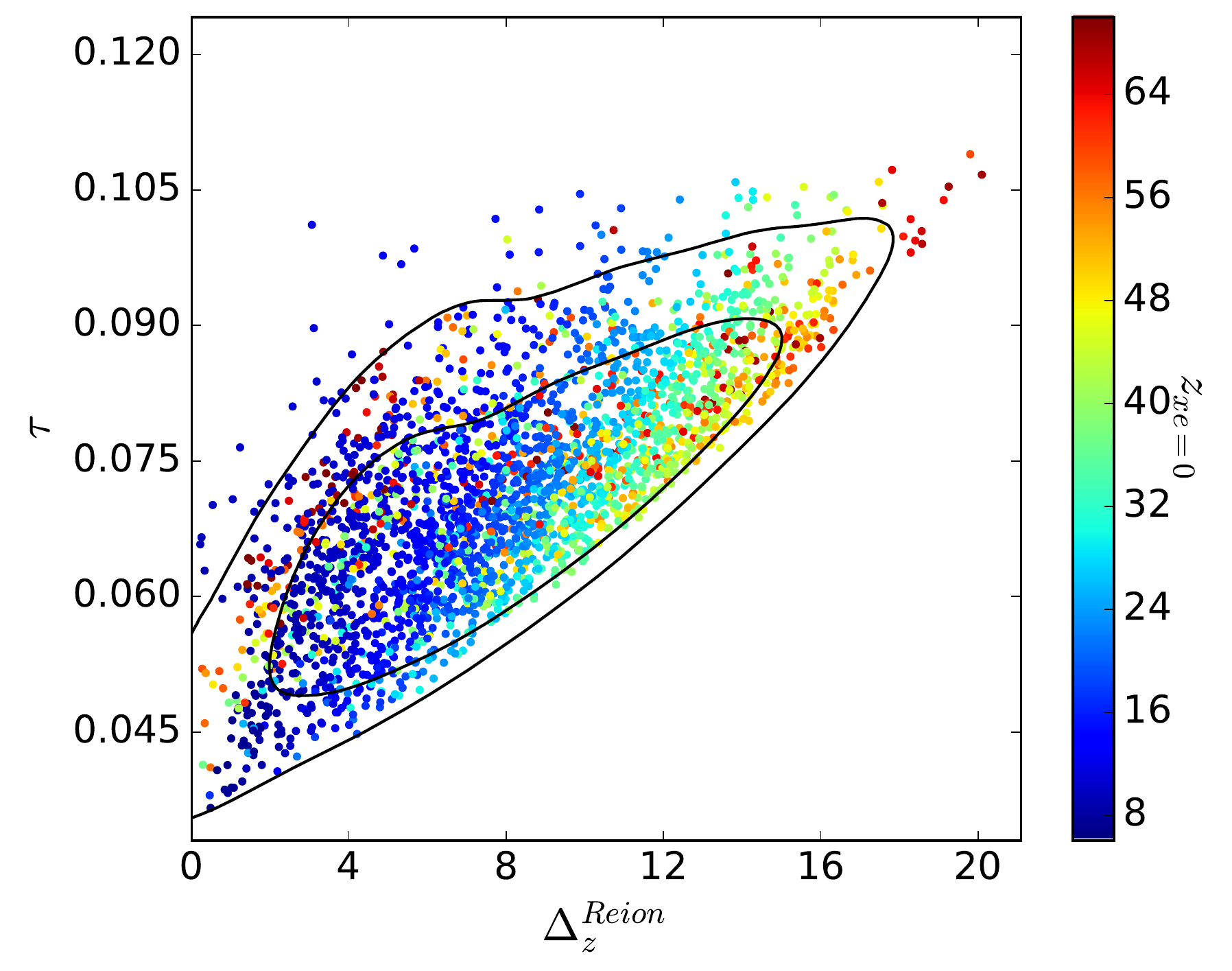}}
\resizebox{205pt}{160pt}{\includegraphics{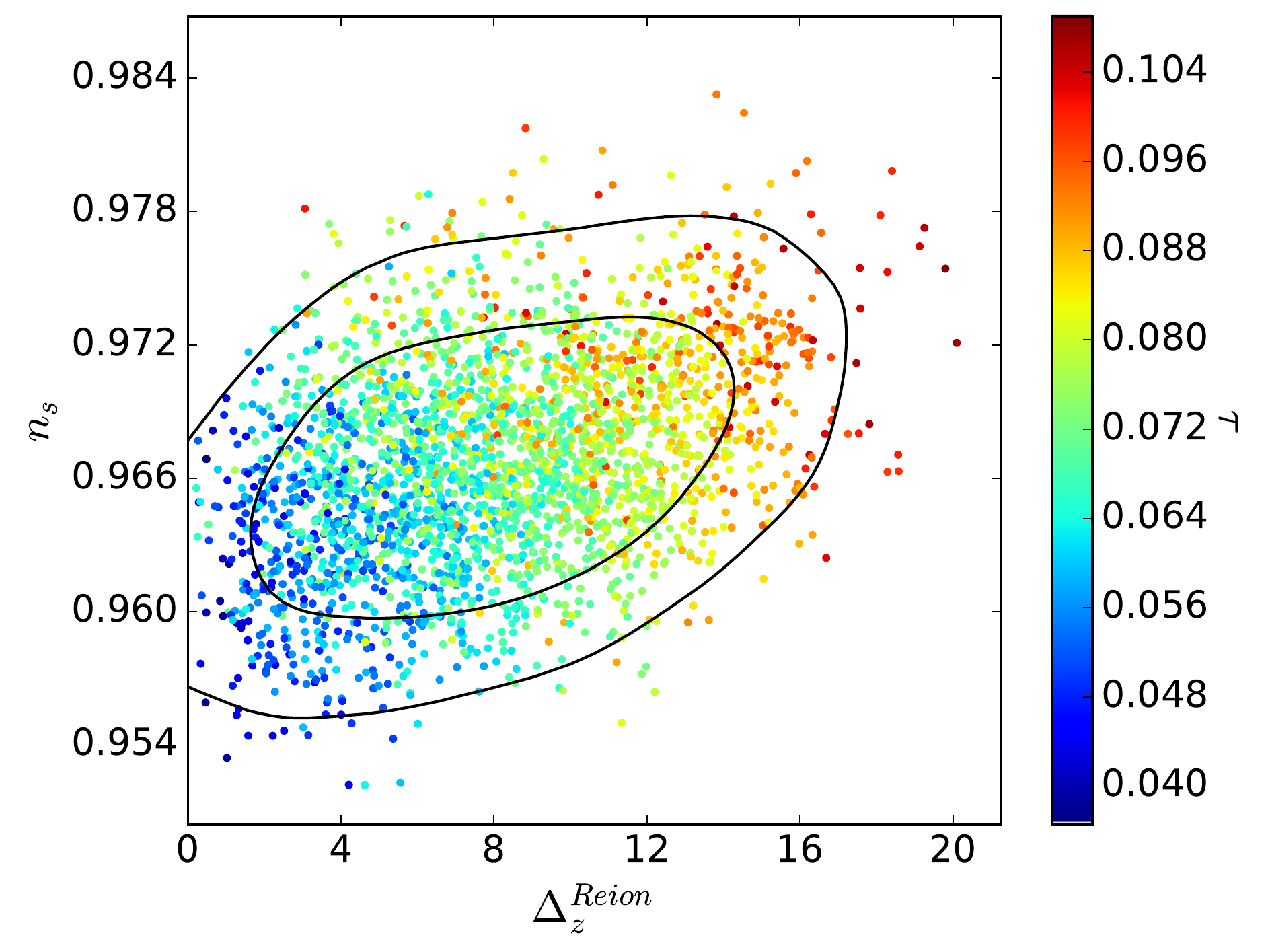}}
\end{center}
\caption{\footnotesize\label{fig:Planck-constraints-LCDM}[Left] Planck-2015 constraints (68\% and 95\% confidence contours are shown) on the duration of reionization and 
the reionization optical depth in the $\Lambda$CDM model. We also plot the redshift in color that marks the beginning of reionization. [Right] Correlation between the 
duration of reionization and scalar spectral index are provided (for $\Lambda$CDM model). The color scatter plots the optical depth that too shows degeneracy as expected.}
\end{figure*}

\begin{figure*}[!htb]
\begin{center} 
\resizebox{205pt}{160pt}{\includegraphics{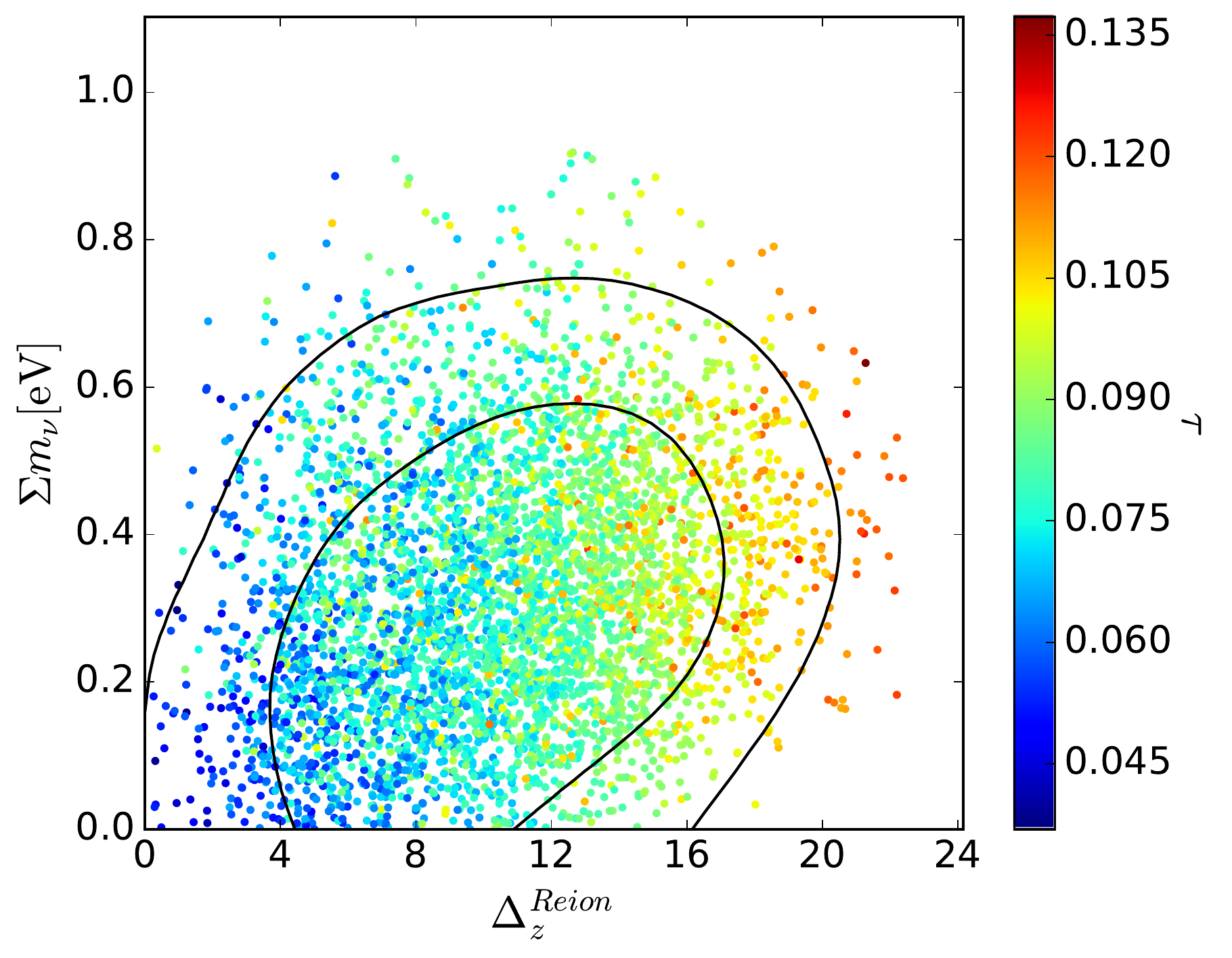}}
\resizebox{205pt}{160pt}{\includegraphics{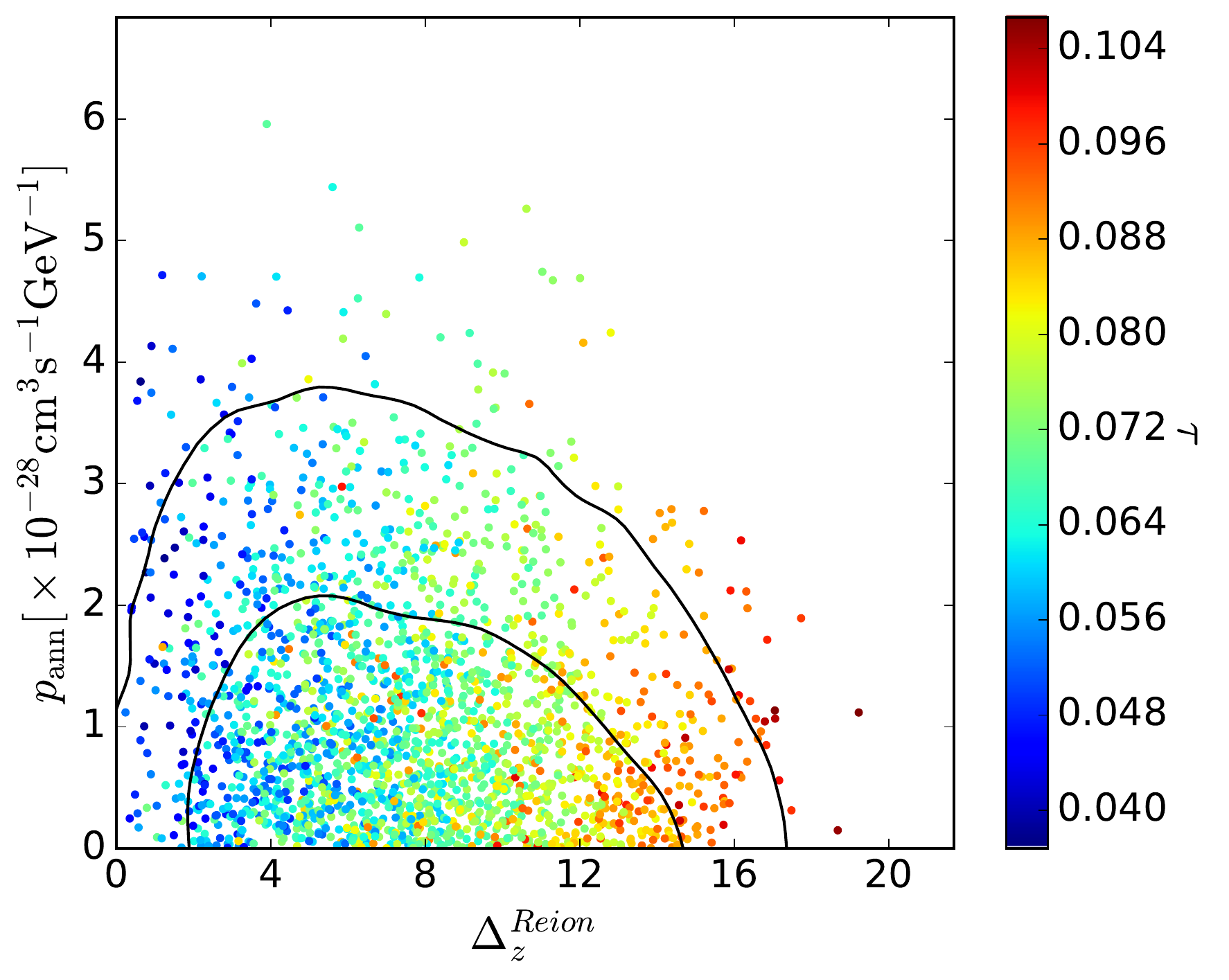}}
\end{center}
\caption{\footnotesize\label{fig:Planck-constraints-beyond-LCDM}Constraints on beyond $\Lambda$CDM Planck baseline model. [Left] When we allow the neutrino mass to vary, 
strong degeneracy with the duration and optical depth emerges in the Planck constraints. 
[Right] When we allow annihilating dark matter, the correlation between the annihilation parameter $p_{\rm ann}$ and the duration of reionization and optical 
depth are shown in the contours and $\tau$ as scatter points.}
\end{figure*}

We provide the current constraints obtained using Planck 2015 data in Fig.~\ref{fig:Planck-constraints-LCDM} (for Planck baseline model) and 
Fig.~\ref{fig:Planck-constraints-beyond-LCDM} (for extended models), in which we plot the cases where reionization optical depth and duration 
of reionization are degenerate with other physical effects. In the left of Fig.~\ref{fig:Planck-constraints-LCDM} we present the results for the
$\Lambda$CDM model with poly-reion. The plot shows the degeneracy between the optical depth and the duration of reionization. With the 
low-$\ell$ joint temperature-polarization Planck 2015 likelihood we find strong correlation between $\tau$ and $\Delta_{z}^{\rm Reion}$. The
beginning of reionization ($z_{xe=0}$) is not constrained well and, as expected, is degenerate with $\tau$ and $\Delta_{z}^{\rm Reion}$ [we have plotted
the points in color scatter]. The value of the optical depth $\tau$ we recovered is in agreement with the Planck 2015 data results \citep{Planck:2015Param}.
We find that extended      
reionization models are allowed but do not find any significance over 2$\sigma$. Top right plot shows the degeneracy between the scalar spectral index ($n_{\rm s}$) 
and $\Delta_{z}^{\rm Reion}$, being $\Delta_{z}^{\rm Reion}$ and $\tau$ degenerate, $n_{\rm s}$ is degenerate with both, as shown by the scatter plot.

The degeneracy between $n_{\rm s}$ and $\tau$ is caused by the fact that the former provides a tilt in the power spectrum and the latter an overall suppression to 
the power spectrum (with a factor $e^{-2\tau}$) at scales smaller than the horizon at the epoch of reionization. The degeneracy is reduced by the temperature data on
large angular scales which provide useful information on the tilt (within the cosmic variance limit).
It is expected that accurate  polarization measurements on large angular scales will be able to further reduce this degeneracy by constraining the 
reionization history and $\tau$.

In Fig.~\ref{fig:Planck-constraints-beyond-LCDM} we present the results for the extension to the $\Lambda$CDM model with poly-reion.
The correlation with the neutrino mass is plotted at the left panel of Fig.~\ref{fig:Planck-constraints-beyond-LCDM}. The total neutrino mass $\Sigma m_{\nu}$ 
changes the expansion history and as a consequence the line of sight integral for the optical depth changes. In order to obtain the same $\tau$ with larger 
total neutrino mass, one needs extended history of reionization to allow for more free electrons along the line of sight. The plot shows that in Planck 2015 
data, $\Delta_{z}^{\rm Reion}$ and $\tau$ are correlated with $\Sigma m_{\nu}$. 

At the right we plot the 2D contour of $p_{\rm ann}$ and the $\Delta_{z}^{\rm Reion}$ with $\tau$ always in scatter points in the case where we 
consider annihilating dark matter. The free electrons generated in this case at high redshifts increase the large and intermediate scale polarization
signal and therefore $p_{\rm ann}$ is degenerate with $\Delta_{z}^{\rm Reion}$ and $\tau$. The small degeneracy occurs since $p_{\rm ann}$ also introduce
damping at the small angular scales which is well constrained by Planck 2015. Therefore we find that a large value of $p_{\rm ann}$ prefers a lower value 
of $\Delta_{z}^{\rm Reion}$ and $\tau$.

\subsection{LiteBIRD and CORE forecasts}

By considering reionization histories allowed by Planck-2015 with a lower value of the average optical
depth as indicated by Ref.~\cite{Adam:2016hgk,Aghanim:2016yuo}, we simulate the fiducial CMB angular power spectra and forecast the constraints by using the noise 
power spectrum from LiteBIRD and CORE-like specifications. In order to compare the constraints between LiteBIRD and CORE in the context of optical depth we
use following three durations of reionization, $\Delta_{z}^{\rm Reion}\sim 5, 8.5$ and $12$.
\begin{figure*}[!htb]
\begin{center} 
\resizebox{140pt}{120pt}{\includegraphics{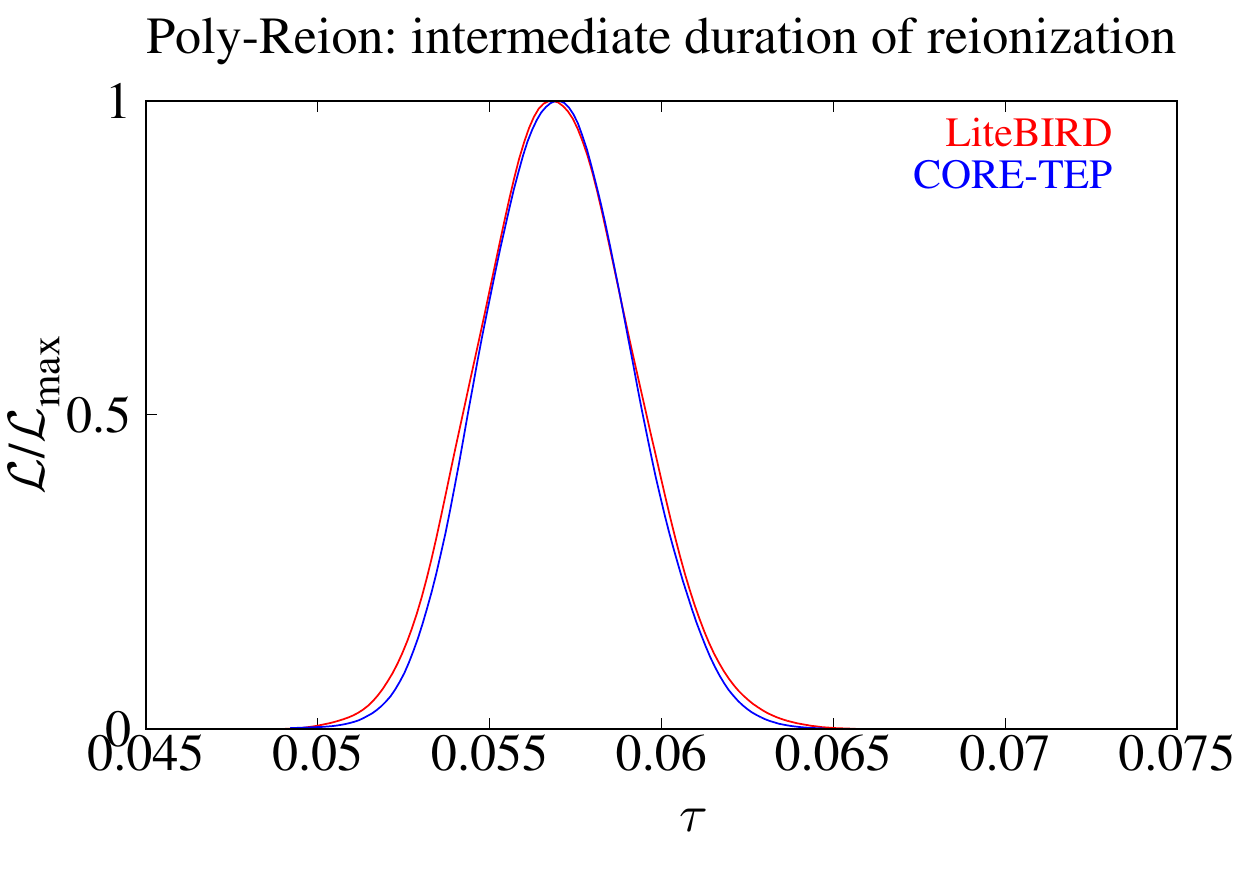}}
\resizebox{140pt}{120pt}{\includegraphics{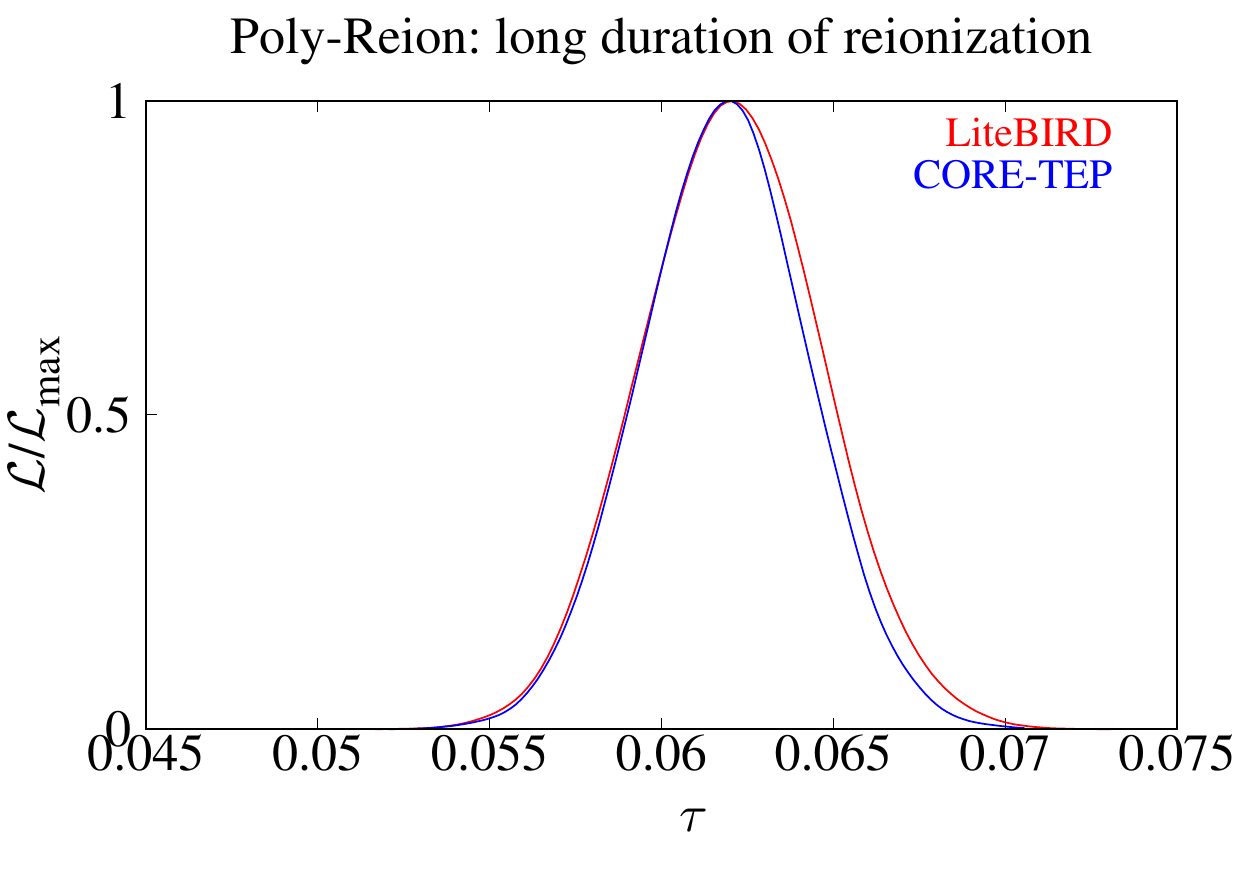}}
\resizebox{140pt}{120pt}{\includegraphics{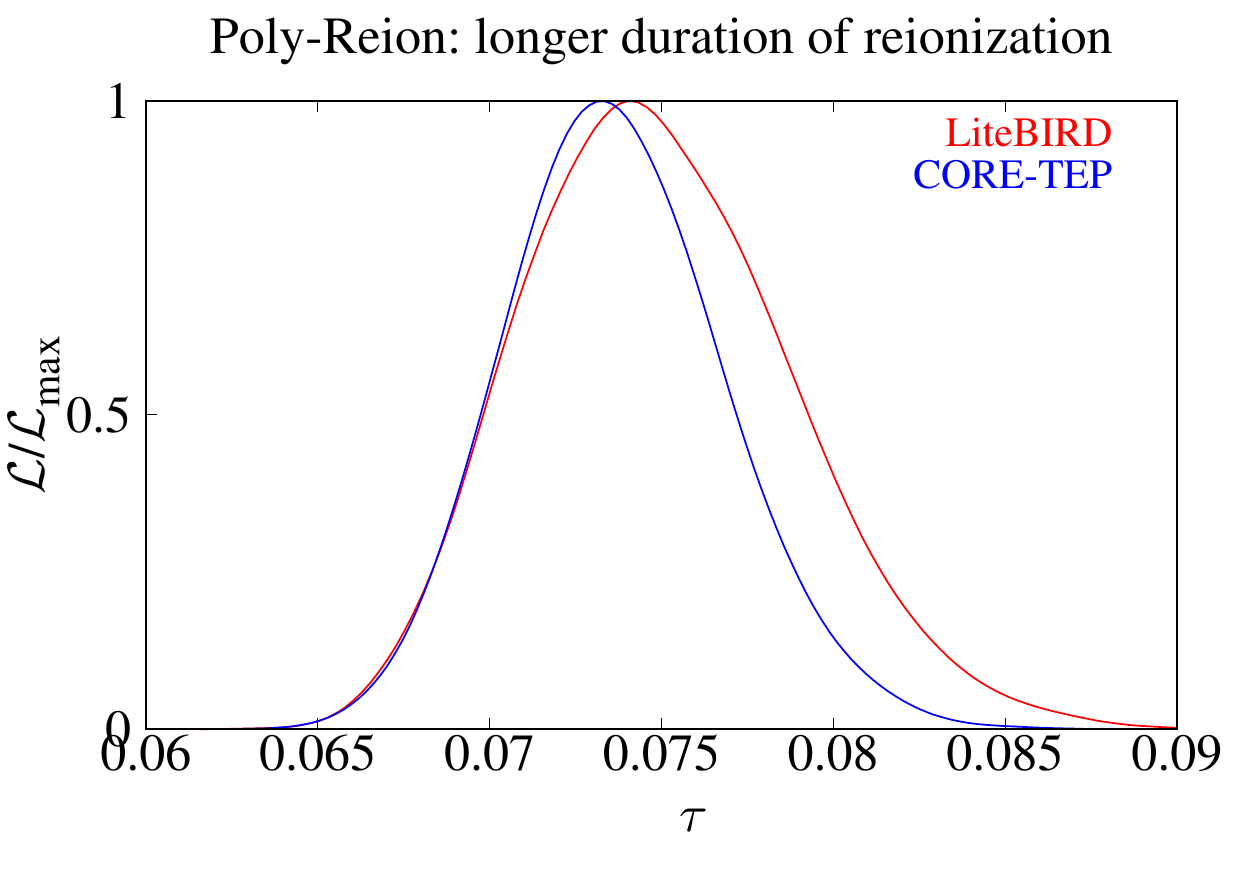}}
\end{center}
\caption{\footnotesize\label{fig:LB-CORE-tau}The constraints on the optical depth compared using proposed LiteBIRD and CORE specifications. Left panel shows
the constraints where we have used the fiducial cosmology from Planck-2015 chains using an intermediate duration of reionization (with $\Delta_z^{\rm Reion}\sim5$).
Middle panel compares the same when we the fiducial with a longer duration ($\Delta_z^{\rm Reion}\sim8.5$). Right panel uses a fiducial with
$\Delta_z^{\rm Reion}\sim12$. Note that as we use higher duration of reionization as fiducial model, we find better constraints from CORE compared to LiteBIRD as
polarization signals at higher multipoles are affected. For better visibility and comparison we have kept the ranges of $\tau$ in the first 2 plots same but in 
the right plot we shifted the covered $\tau$ range to match the fiducial used.}
\end{figure*}

\begin{figure*}[!htb]
\begin{center} 
\resizebox{205pt}{205pt}{\includegraphics{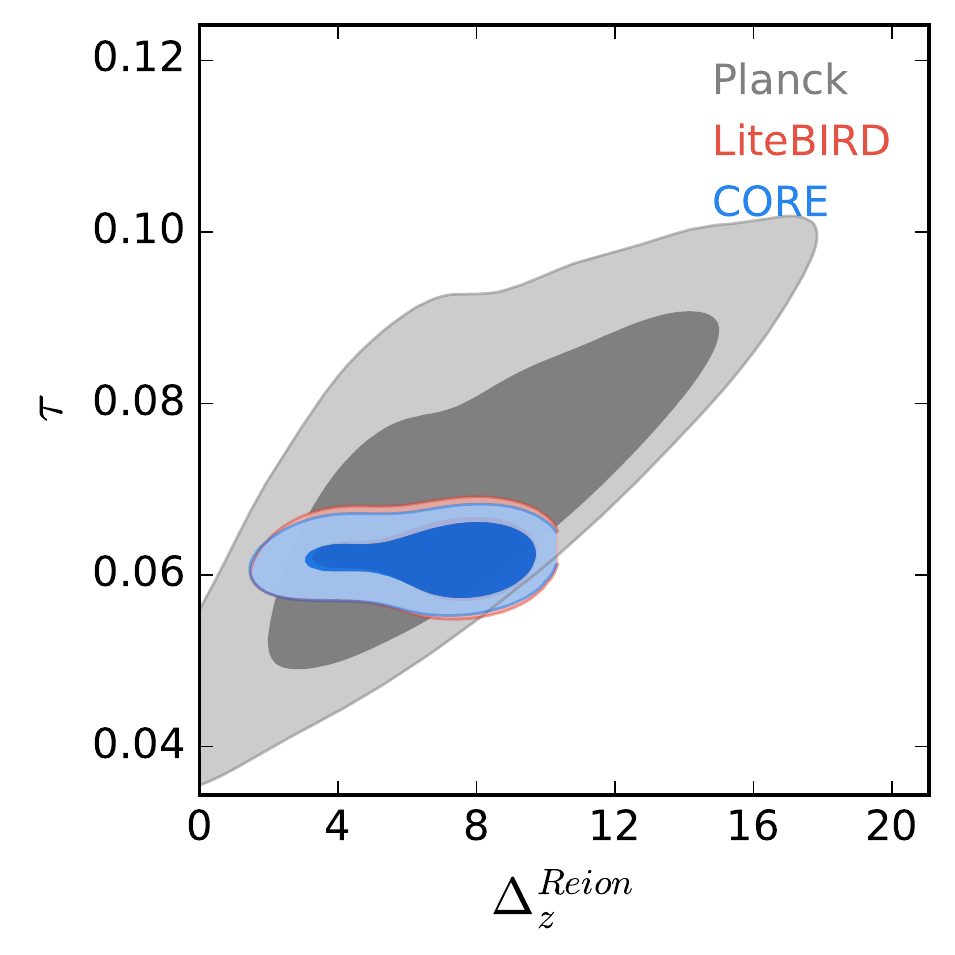}}
\resizebox{205pt}{205pt}{\includegraphics{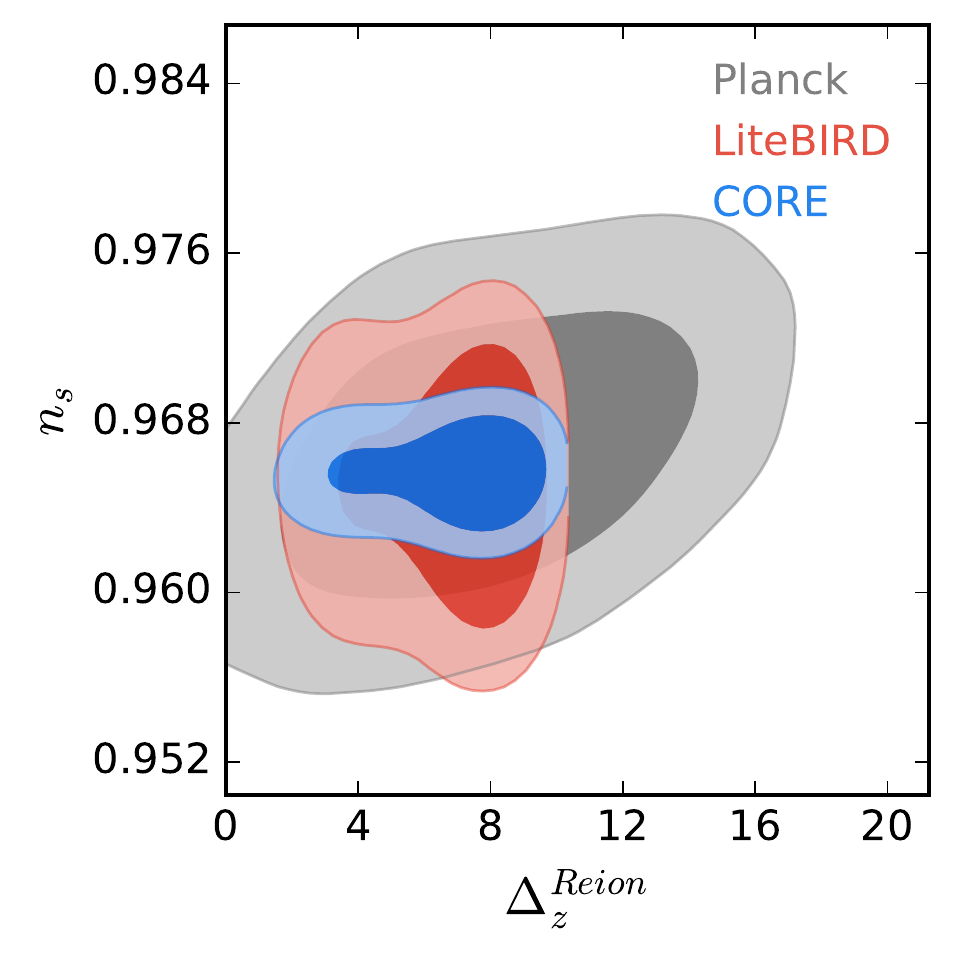}}
\end{center}
\caption{\footnotesize\label{fig:Pl-LB-CORE-LCDM}Comparison between Planck constraints and projected LiteBIRD and CORE constraints using a fiducial cosmology
in agreement with Planck-2015 with an extended reionization history ($\Delta_z^{\rm Reion}\sim8.5$) and $\tau\sim0.06$, in agreement with the latest Planck-2016 
results on large angular scales in polarization. [Left] Comparison in the context of degeneracies between the duration and optical depth of reionization in
the $\Lambda$CDM model. [Right] Degeneracy between the reionization duration and scalar spectral index.}
\end{figure*}

\begin{figure*}[!htb]
\begin{center} 
\resizebox{205pt}{205pt}{\includegraphics{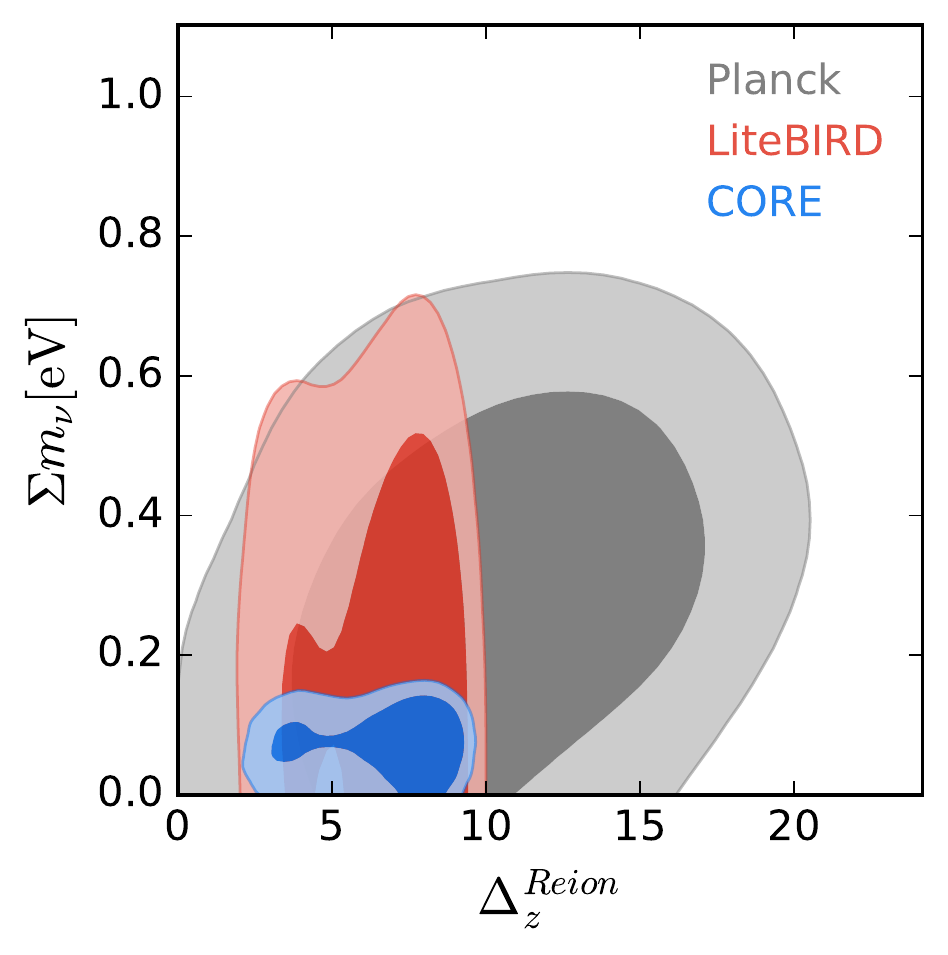}}
\resizebox{205pt}{205pt}{\includegraphics{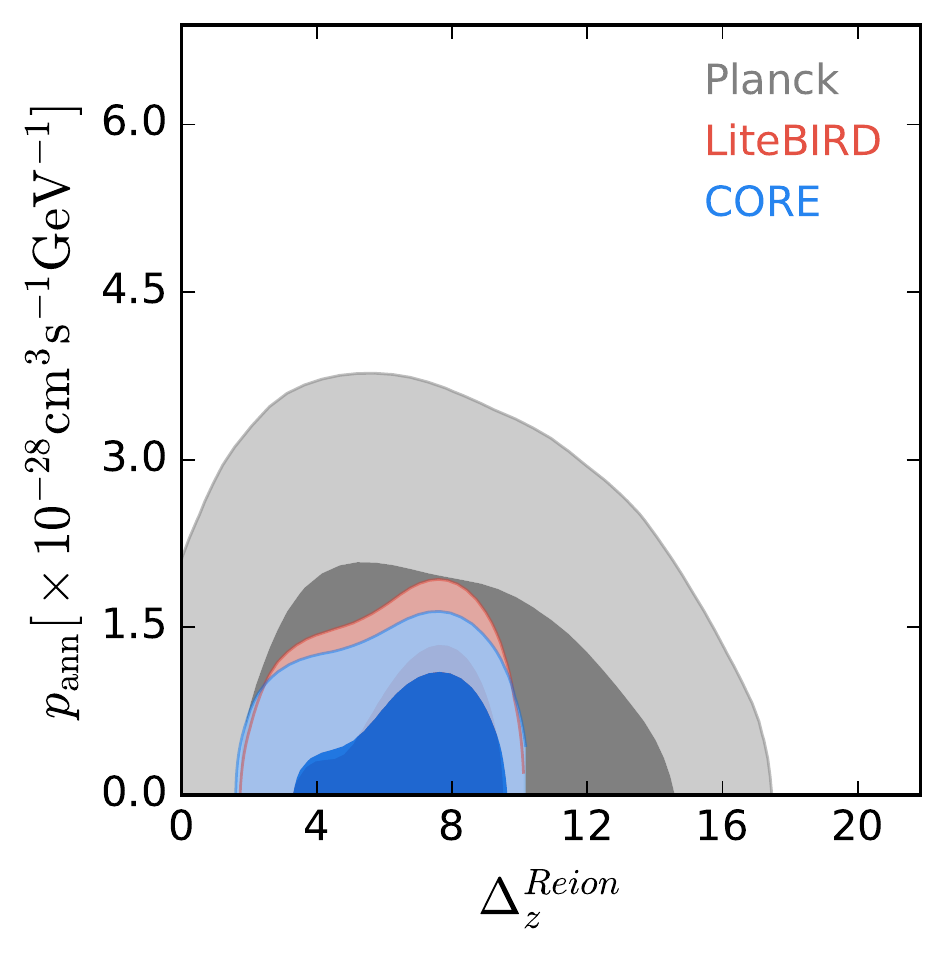}}
\end{center}
\caption{\footnotesize\label{fig:Pl-LB-CORE-beyond-LCDM}Comparison of present and projected constraints as in Fig.~\ref{fig:Pl-LB-CORE-LCDM} for models beyond
Planck baseline analysis. [Left] Comparison when neutrino mass is allowed to vary in a degenerate hierarchy. [Right] Existing and future possible correlation 
between the dark matter annihilation parameter and the duration of reionization.}
\end{figure*}

\begin{figure*}[!htb]
\begin{center} 
\resizebox{300pt}{250pt}{\includegraphics{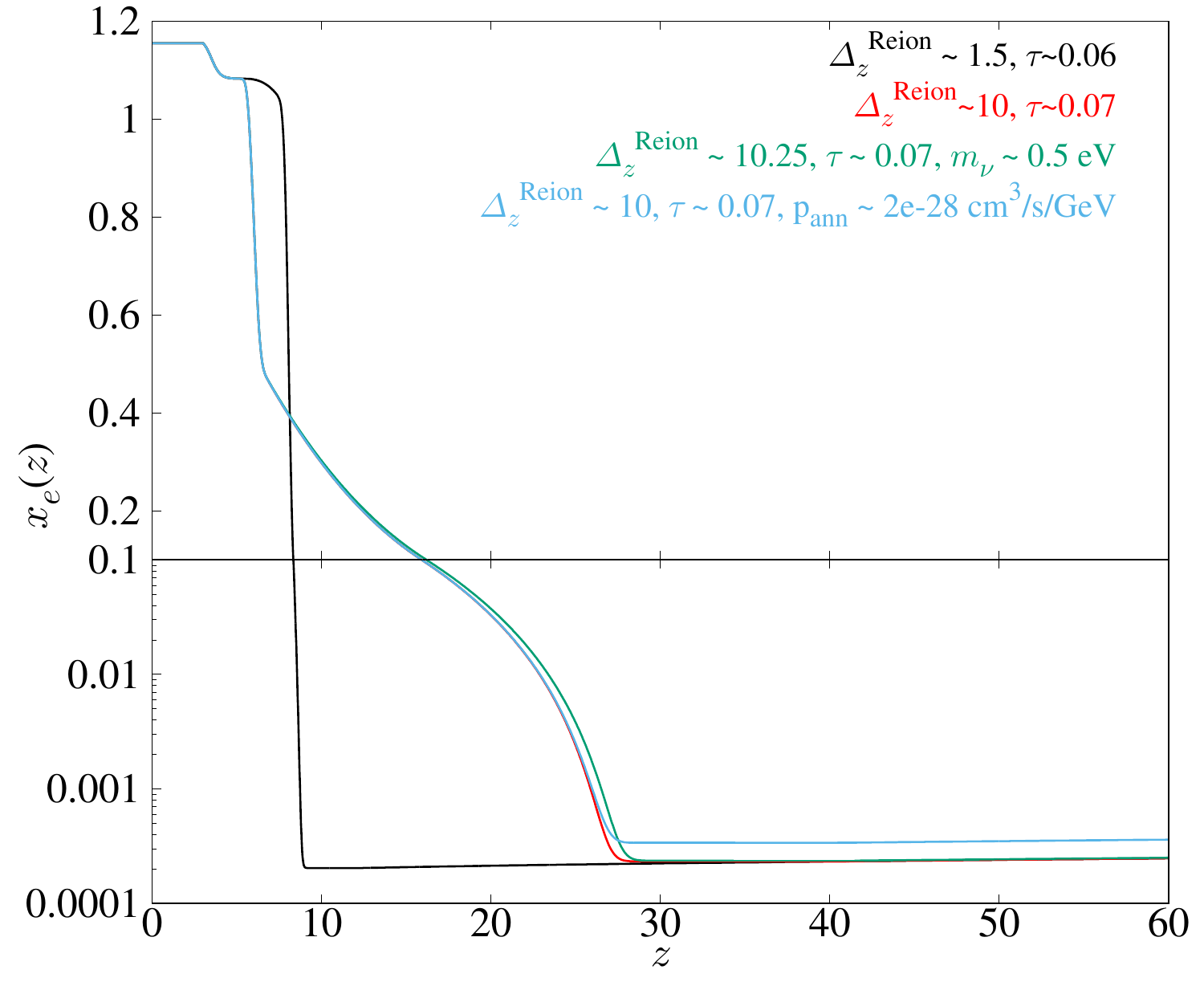}}
\end{center}
\caption{\footnotesize\label{fig:xe}History of reionization plotted for 4 cases. The black like represents the electron fraction for a 
nearly instantaneous reionization with a $\Delta_{z}^{\rm Reion}=1.5$. In order to demonstrate the comparison in the next plots we will 
use this history as baseline. The extended reionization we consider here is plotted for $\Delta_{z}^{\rm Reion}=9.5$ and with a value 
of optical depth of 0.07. Keeping the same optical depth if we assume the total mass of neutrino to be $0.5{\rm eV}$, then due to the 
change in the expansion history we need a longer duration of reionization. If dark matter annihilates with a $p_{\rm ann}=2\times10^{-28} {\rm cm^3/s/GeV}$, 
it increases the number of free electrons at higher redshifts compared to $\Lambda$CDM model as shown in the plots.} 
\end{figure*}

Note that we use these durations from the Planck-2015 chains that agree well with the data. We have used the corresponding values of other cosmological 
parameters from the chains directly. For these three values of $\Delta_{z}^{\rm Reion}$ the corresponding optical depths are 0.057, 0.06 and 0.07 
respectively. Note that for the first two fiducials the optical depths are nearly indistinguishable even with future surveys (as in CORE-like $\delta\tau\sim0.002$ 
at 68\% C.L.~\cite{core:cosmoparam}). 

The constraints on $\tau$ are shown in Fig.~\ref{fig:LB-CORE-tau}. The left to right panels are provided in the increasing order of duration of reionization. 
We label them as intermediate, long and longer duration of reionization. We find that for intermediate case, the projected constraints on optical depth is very
similar in both proposed missions. For long and longer durations of reionization, CORE performs better as in these two scenarios of extended reionization the
polarization power spectrum is enhanced at higher multipoles where CORE-like performs better then LiteBIRD due to its smaller beam. For the longer duration 
(plot at the right) we find CORE specification can provide 20\% improvement in the constraint at $2\sigma$ compared to LiteBIRD. Therefore, for extended reionization
scenarios, we need CORE-like mission in order to obtain the best constraint. 

\begin{figure*}[!htb]
\begin{center} 
\resizebox{430pt}{220pt}{\includegraphics{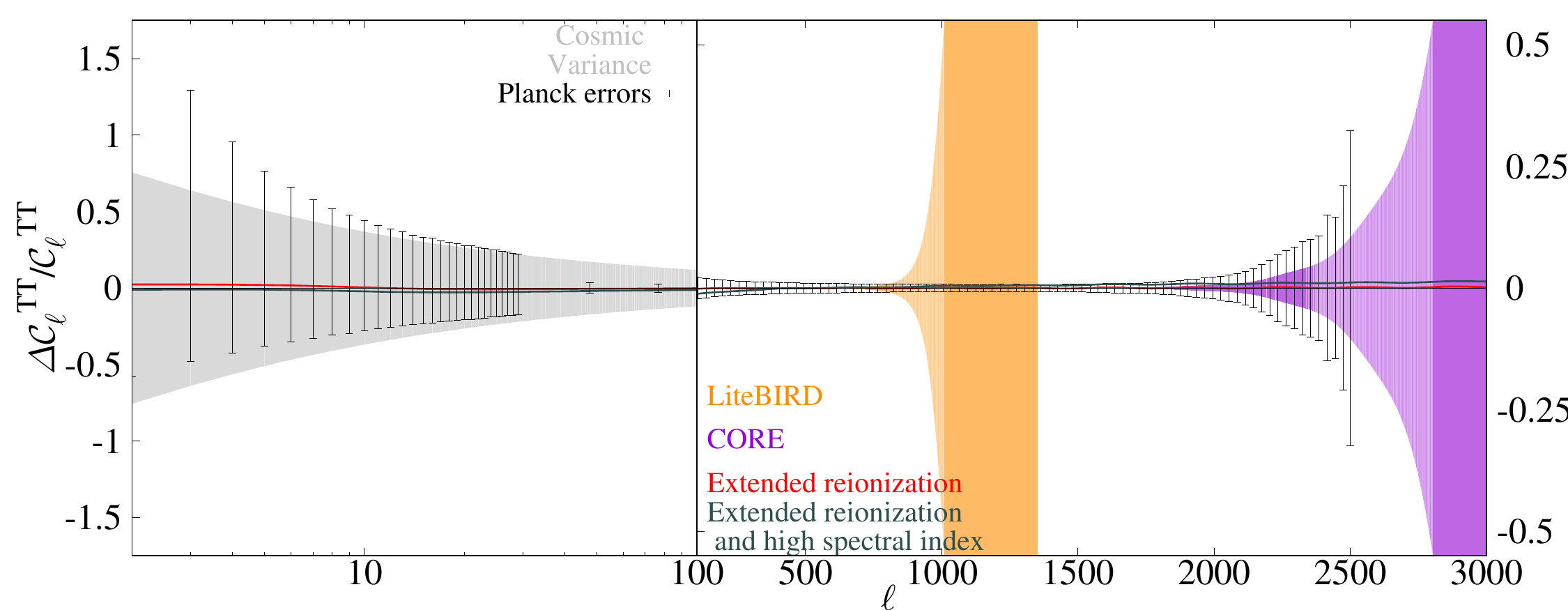}}

\resizebox{430pt}{220pt}
{\includegraphics{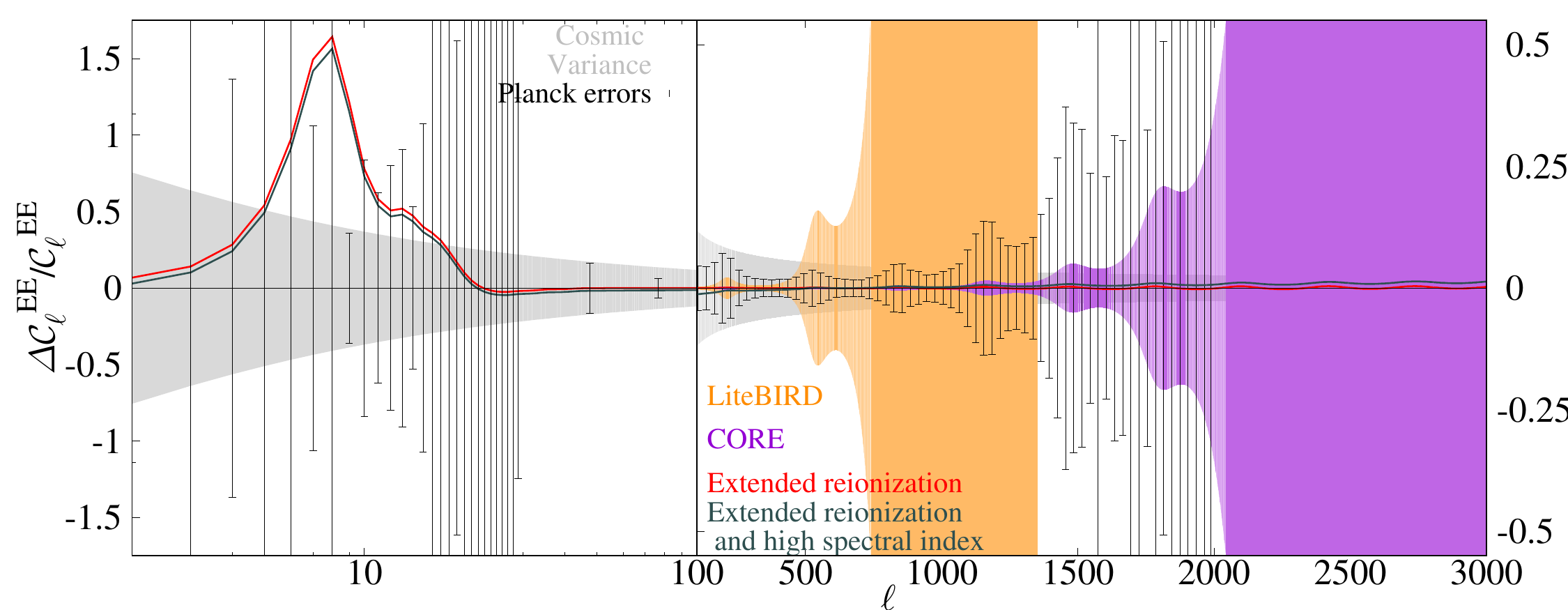}}
\end{center}
\caption{\footnotesize\label{fig:delta-cl-ns}Fractional change in the temperature and polarization anisotropy power spectrum. 
We plot the changes {\it w.r.t.} the baseline mentioned in Fig.~\ref{fig:xe}. In particular, we plot the differences when extended 
reionization model with a $\Delta_{z}^{\rm Reion}=9.5$ is considered with same spectral index and with a spectral index higher 
($n_{\rm s}=0.972$) than the baseline ($n_{\rm}=0.965$) is considered. Planck error-bars (scaled with the power spectrum) are 
plotted unbinned for low multipoles and binned for high multipoles. Cosmic variance uncertainties with LiteBIRD and CORE proposed
noise are provided in shaded regions.}
\end{figure*}

\begin{figure*}[!htb]
\begin{center} 
\resizebox{430pt}{220pt}{\includegraphics{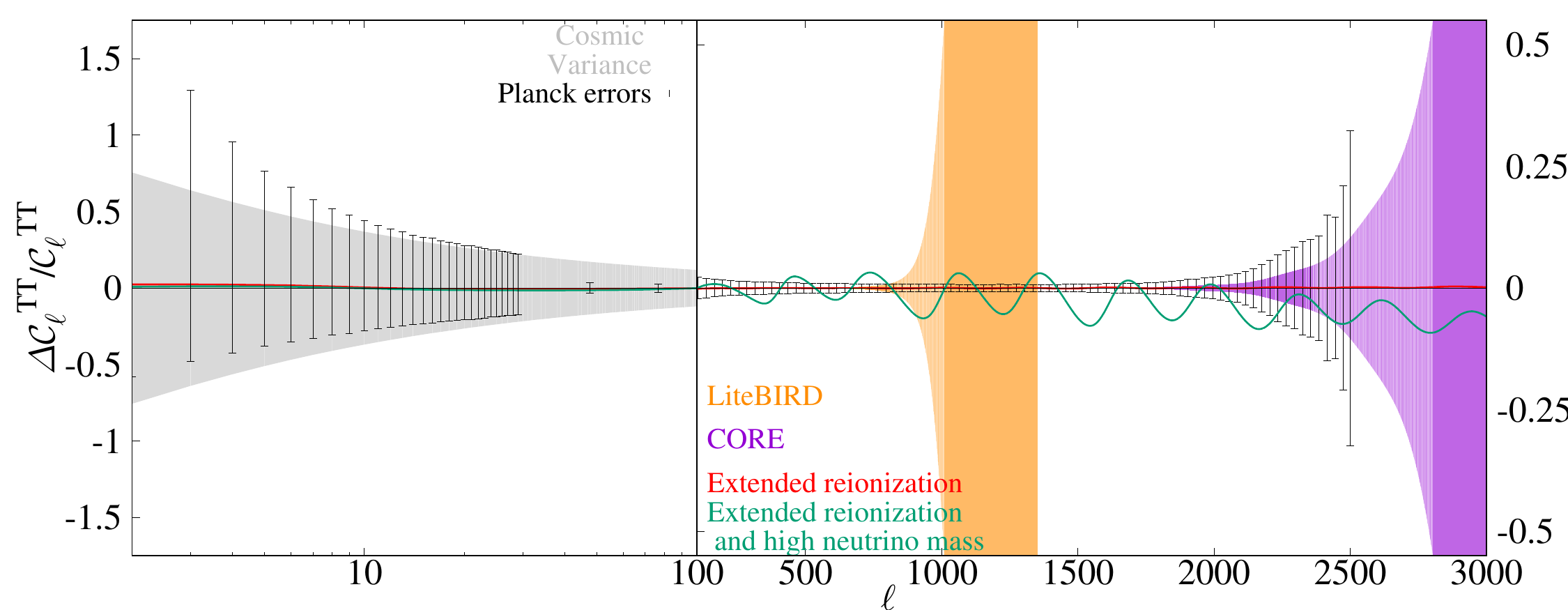}}

\resizebox{430pt}{220pt}
{\includegraphics{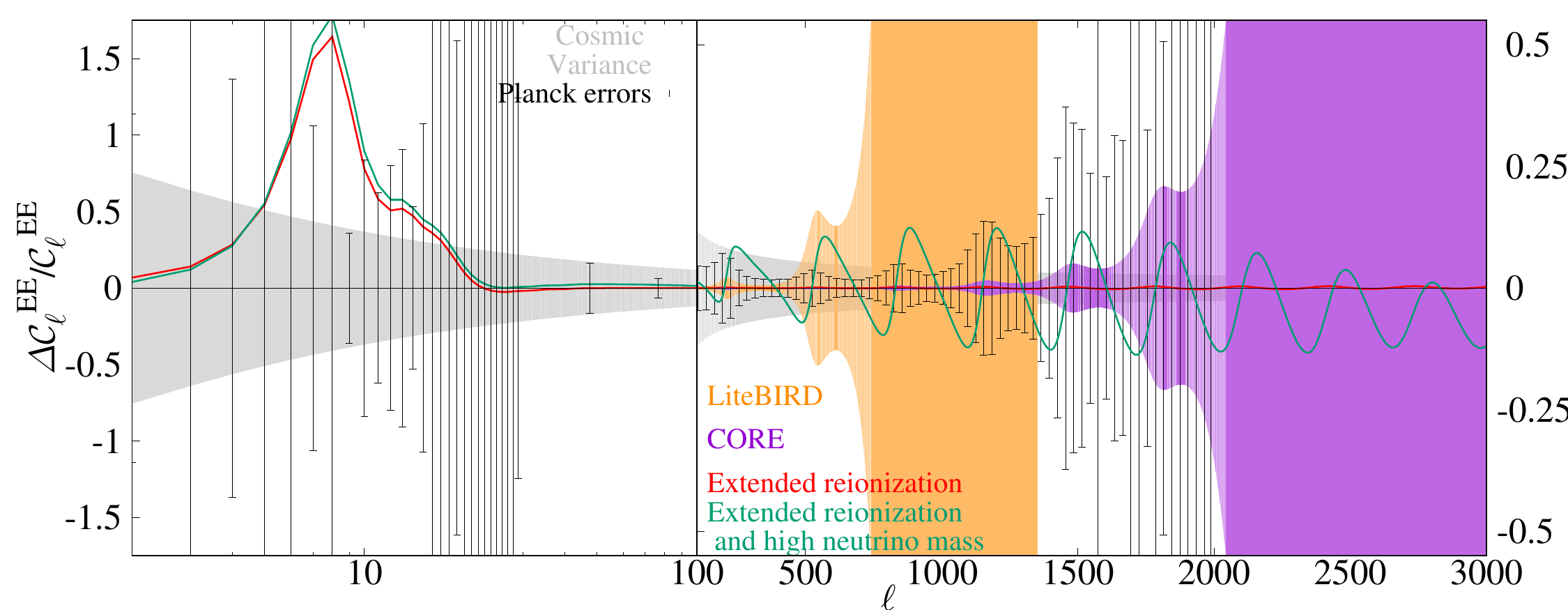}}
\end{center}
\caption{\footnotesize\label{fig:delta-cl-mnu}Same as in Fig.~\ref{fig:delta-cl-ns} but here we compare the differences when the 
neutrino mass is assumed to be $0.5{\rm eV}$ keeping all other parameter fixed to the extended reionization scenarios. Note that 
the change in the distance measure here requires longer history of reionization (see, Fig.~\ref{fig:xe}) for the same optical depth 
and therefore it increases the polarization signal at large angular scales.}
\end{figure*}

\begin{figure*}[!htb]
\begin{center} 
\resizebox{430pt}{220pt}{\includegraphics{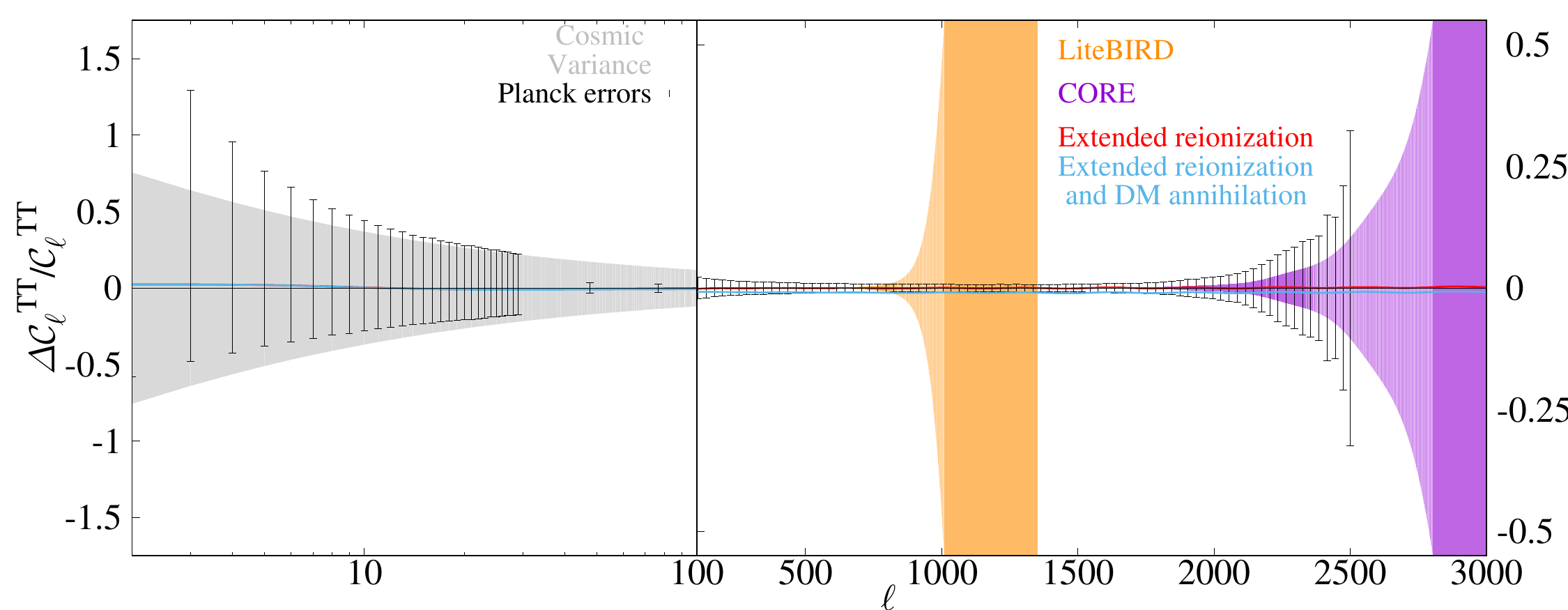}}

\resizebox{430pt}{220pt}
{\includegraphics{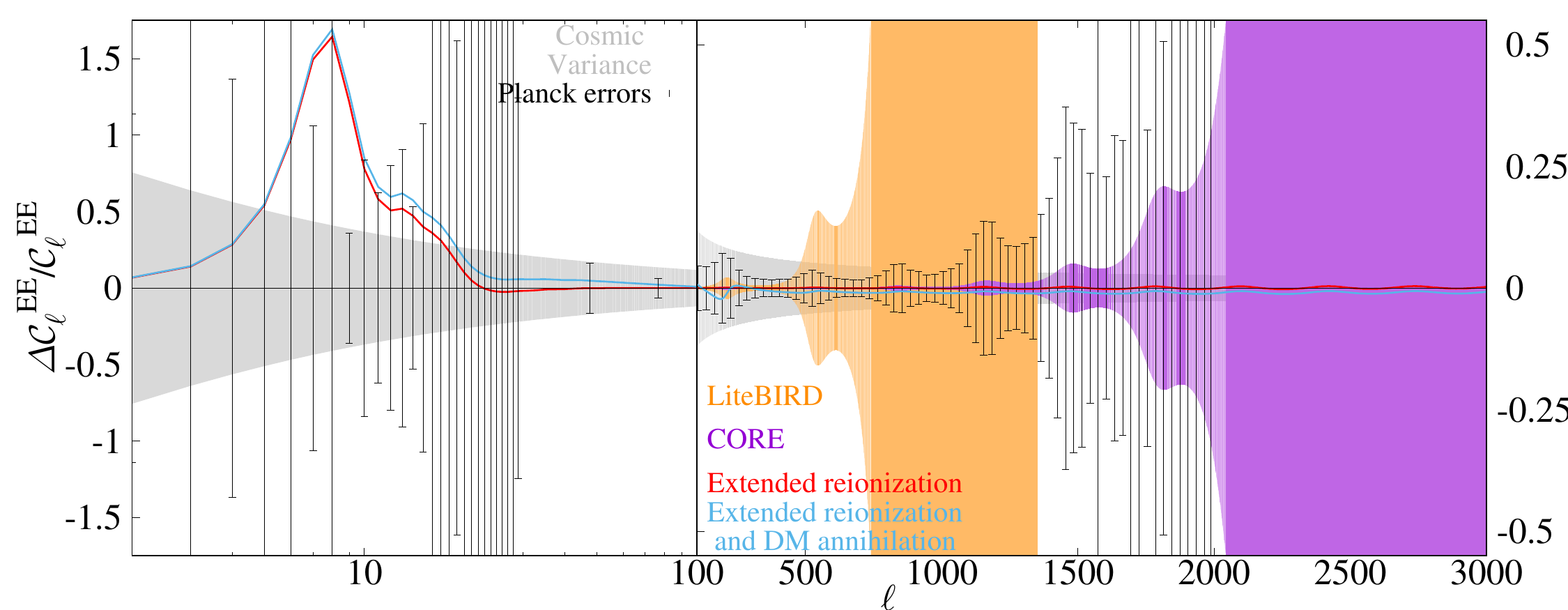}}
\end{center}
\caption{\footnotesize\label{fig:delta-cl-dmann}Same as in Fig.~\ref{fig:delta-cl-ns} but here we compare the differences when 
the dark matter annihilation is allowed with $p_{\rm ann}=2\times10^{-28} {\rm cm^3/s/GeV}$. Higher fraction of free electrons 
here too increases the polarization signal through Thomson scattering. Corresponding free electron function is plotted in Fig.~\ref{fig:xe}.}
\end{figure*}

We now provide the forecasts on the degeneracies between reionization and other physical processes with LiteBIRD and CORE. In all the cases we select the fiducial from $\Delta_{z}^{\rm Reion}\sim8.5$. 
In Figs.~\ref{fig:Pl-LB-CORE-LCDM} and~\ref{fig:Pl-LB-CORE-beyond-LCDM} we provide the Planck constraints that have already been discussed in Figs.~\ref{fig:Planck-constraints-LCDM} 
and~\ref{fig:Planck-constraints-beyond-LCDM} in grey contours as a comparison. 
Over that we provide the predicted LiteBIRD constraints in red and the CORE constraints in blue.

The results show that for both future experiments there will not be any correlation between $\tau$ and $\Delta_{z}^{\rm Reion}$. We find the constraints to be nearly 
the same in both the cases since most of the improvement just comes from the large scale polarization data and both the surveys provide cosmic variance limited 
measurements. The right plot in Fig.~\ref{fig:Pl-LB-CORE-LCDM} of provides the correlation between $n_{\rm s}$ and $\Delta_{z}^{\rm Reion}$. We note that with future 
experiments the degeneracy is broken. However, as expected \citep{core:cosmoparam,core:inf}, LiteBIRD and CORE proposals provide different performances for the 
constraint on $n_{\rm s}$. 

On the left of Fig.~\ref{fig:Pl-LB-CORE-beyond-LCDM} we witness similar breaking of degeneracies between the neutrino mass and the reionization history. 
Note that our conclusions on the removal of degeneracy between the duration of reionization and the total neutrino mass agree with the analysis
carried in~\cite{core:cosmoparam}, which however used a different parametrization of an extended reionization process. Even with neutrino mass 
treated as free parameter, we find that LiteBIRD and CORE are performing similar in constraining the $\Delta_{z}^{\rm Reion}$. Since CORE is 
constraining $\Sigma m_{\nu}$ nearly an order of magnitude better than LiteBIRD it is able to reduce the residual degeneracies with neutrino mass and 
other parameters and provides a slightly better constraint on reionization history. We provide the 95\% upper limit on the total neutrino mass in
Table~\ref{tab:ul}. If we consider a longer duration of the reionization, as we have shown for the $\Lambda$CDM, CORE can be in a more advantageous 
position.

For dark matter annihilation, at the right of Fig.~\ref{fig:Pl-LB-CORE-beyond-LCDM} we plot the constraints on $p_{\rm ann}$ and 
$\Delta_{z}^{\rm Reion}$. We find that at 95\% level, the degeneracy is lifted in both cases. We note that $p_{\rm ann}$ constraint 
is similar in LiteBIRD and CORE. Since dark matter annihilation increases the free electron fraction by heating up the IGM, we have 
an excess of free electrons before the reionization by high energetic sources begin. These electrons change the polarization signal 
at intermediate scales and therefore can be constrained with similar upper limits as in LiteBIRD and CORE. Since the annihilation also 
changes the epoch of recombination, the peak positions gets changed by small amount and there CORE can be more powerful than LiteBIRD 
in that intermediate scale regime. We find that $p_{\rm ann}$ is constrained nearly 2.5 times better compared to Planck 2015 as shown in Table~\ref{tab:ul}.

\begin{table*}
\begin{scriptsize}
\begin{center}
\tabcolsep=0.05cm
  \hspace*{0.0cm}\begin{tabular}
  {c|c|c|c } \hline 
  
  Data & Planck-TEP & LiteBIRD-TE & CORE-TEP  \\ \hline 

  $\Sigma m_{\nu}~[{\rm eV}]$ 	& 0.63 &0.58 &0.14 \\  \hline

  $p_{\rm ann}[\times 10^{-28} {\rm cm^3/s/GeV}]$ 	&3 &1.5 &1.3 \\\hline

  \end{tabular}
  \end{center}
\caption{~\label{tab:ul}95\% upper limits on neutrino mass and the dark matter annihilation parameter from Planck 2015 and forecasts using 
LiteBIRD and CORE specifications. In both the cases poly-reion is used as the history of reionization.}
\end{scriptsize}
\end{table*}

In order to provide a visual render of the degeneracies we show the ionization fraction for a few reionization histories in Fig.~\ref{fig:xe}.
In black we plot a nearly instantaneous reionization history with $\Delta_{z}^{\rm Reion}\sim1.5$ that leads to $\tau\sim0.06$. We consider 
one extended reionization history with $\Delta_{z}^{\rm Reion}\sim10,~\tau\sim0.06$ in red. Note that in this latter case although reionization 
is 10\% complete by $z\sim16$, the electron fraction merges to its value after recombination at around $z=30$. The green line plots reionization
history for a $\Sigma m_{\nu}\sim0.5~{\rm eV}$. Keeping the other parameters fixed if we increase the neutrino mass, it changes the expansion 
history and therefore in order to obtain the same integrated optical depth, we need longer period of reionization. Whereas for an optical depth
of 0.07 we find the $\Delta_{z}^{\rm Reion}$ changes by 0.25, for higher optical depth the difference becomes substantial and Planck captures 
this degeneracy. The blue line plots the reionization history for the same parameters in the dark matter annihilation scenario. Note that the
electron fraction here merges to a higher value at the beginning of reionization, that represents the excess electrons generated in this case.   

In order to visualize the changes on the power spectra induced by these models, we plot the relative differences in the angular power spectrum of
temperature and polarization anisotropies in Figs.~\ref{fig:delta-cl-ns},~\ref{fig:delta-cl-mnu} and~\ref{fig:delta-cl-dmann} respectively for 
$\Lambda$CDM Planck baseline, neutrino mass and dark matter annihilation cases. In top panels we plot the differences in the temperature and in the bottom we plot 
the differences in the E-mode polarization. Up to a multipole of 100, we use the logarithmic scale and beyond that the linear scale is used. 
The differences are plotted {\it w.r.t.} the nearly instantaneous reionization with $\Delta_{z}^{\rm Reion}\sim1.5$ in Fig.~\ref{fig:xe}. 
Note that the color codes in Fig.~\ref{fig:xe} are maintained in all the plots. Planck error bars are plotted in black, unbinned for 
the low-$\ell$ part and binned for high-$\ell$. Cosmic variance is plotted in the grey band and the error bands for LiteBIRD and CORE are in orange and purple respectively. 

In all the cases, it is evident that the amplification in polarization spectrum due to extended reionization is still within Planck 2015 polarization uncertainties but surely will be significantly detected with any cosmic variance limited future observation.
Concerning the spectral index we see that the changes on small angular scales may be in the CORE target. A small increase, still fully compatible with Planck data, from $n_{\rm s}=0.965$ to $n_{\rm s}=0.972$, induces a blue tilt lowering the low-$\ell$ polarization signal. These tilt effect is within the noise of the LiteBIRD experiment
which has only the large angular scale lever to use, whereas for CORE the changes both in temperature and polarization will be clearly distinguishable between these two indices. 

For neutrino mass, the effect of the longer duration of reionization is clearly visible at large scale polarization. However, whether we can constrain neutrino mass, that can be 
answered from the changes in high-$\ell$. Due to changes in the expansion history, peak positions are shifted which result in oscillations as residuals. 
Both temperature and polarization sensitivity from CORE type survey will be able to rule out $0.5{\rm eV}$ total neutrino mass with high significance. Note that it is presently within $1-2\sigma$ in Planck.

Excess polarization signal at large scales for dark matter annihilation is plotted in blue line. While the value of $p_{\rm ann}\sim2\times10^{-28}~{\rm cm^3/s/GeV}$ used to generate this plot is outside 68\% confidence interval in Planck, it will be highly disfavored by both LiteBIRD and CORE, as the changes in the intermediate scale temperature and polarization anisotropy can be distinguished by both. 

\section{Conclusion}~\label{sec:conclusion}

In this paper we have discussed the effect of extended reionization in the light of Planck 2015
data. We have restricted ourselves to minimal monotonic physical histories of reionization as 
parametrized by two additional extra parameters namely, an intermediate position in redshift and the free electron fraction at that redshift. 
Using this simple extension of poly-reion model, 
we have studied degeneracies which involve the duration of reionization in $\Lambda$CDM model and in two extended cosmological models, 
{\it i.e.} when the total neutrino mass is allowed to vary and in the case of self-annihilating dark matter, respectively.

In all the models we have discussed the level of degeneracies present with the publicly available Planck 2015 data
between the reionization and other physical processes in standard baseline and a couple of beyond standard cosmological models. In the model considered we have found degeneracies of the duration 
of reionization with the scalar spectral index, with the total neutrino mass and, to a smaller extent, with the dark matter annihilation parameter $p_{\rm ann}$. 

We have then discussed how future CMB experiments dedicated to polarization  
could remove these degeneracies for three different fiducial values of the duration of reionization. We take 
as examples the specifications of LiteBIRD and CORE proposals, which will provide a cosmic variance limited 
measurement of E-mode polarization at large angular scales.
Our study shows that LiteBIRD and CORE are able to remove the degeneracies of the parameters of the 
reionization process with other physical processes and to provide similar
constraints on the reionization epoch, except for the case of 
a long duration of the reionization epoch in which a CORE-like higher angular resolution plays a non-negligible role.
Of course, a CORE-like experiment alone would provide tighter constraints on the parameters other than the reionization history.

\section*{Acknowledgments}
The authors would like to acknowledge the use of APC cluster (\href{https://www.apc.univ-paris7.fr/FACeWiki/pmwiki.php?n=Apc-cluster.Apc-cluster}{https://www.apc.univ-paris7.fr/FACeWiki/pmwiki.php?n=Apc-cluster.Apc-cluster}).
DKH would like to thank Suvodip Mukherjee and Arindam Chatterjee for useful discussions.
GFS acknowledge Laboratoire APC-PCCP, Universit\'e Paris Diderot and Sorbonne Paris Cit\'e (DXCACHEXGS)
and also the financial support of the UnivEarthS Labex program at Sorbonne Paris Cit\'e (ANR-10-LABX-0023 and ANR-11-IDEX-0005-02).
DP and FF acknowledge financial support by ASI Grant 2016-24-H.0 and 
partial financial support by the ASI/INAF Agreement I/072/09/0 for the Planck LFI Activity of Phase E2.

\end{document}